\documentclass[amsmath, amssymb, english, aps,  reprint,floatfix,
superscriptaddress,showpacs,prb]{revtex4-1}
\usepackage[english]{babel} \usepackage{graphics} \usepackage{graphicx}
\usepackage{epsfig} \usepackage{verbatim}

\usepackage[usenames,dvipsnames]{color}
\usepackage{color}

\begin{document}

\title{Spin-orbit coupling in three-orbital Kanamori impurity model
and its relevance for transition-metal oxides }

\author{Alen~Horvat} 
\affiliation{Jo\v{z}ef Stefan Institute, Jamova~39, Ljubljana, Slovenia} 
\author{Rok~\v{Z}itko} 
\affiliation{Jo\v{z}ef Stefan Institute, Jamova~39, Ljubljana, Slovenia}
\affiliation{Faculty of Mathematics and Physics, University of Ljubljana, Jadranska 19, Ljubljana, Slovenia}
\author{Jernej~Mravlje}
\affiliation{Jo\v{z}ef Stefan Institute, Jamova~39, Ljubljana, Slovenia}

\def\Tk{$T_\mathrm{K}$}
\def\Tkorb{$T_\mathrm{K}^\mathrm{orb}$}
\def\Tksp{$T_\mathrm{K}^\mathrm{spin}$}
\def\Tkz{$T_\mathrm{K}^{0}$}
\def\mTk{T_\mathrm{K}}
\def\Jh{$J_{\mathrm{H}}$}
\def\Jhm{J_{\mathrm{H}}}
\def\mJh{J_{\mathrm{H}}}
\def\t2g{$\mathrm{t}_{2g}$}
\def\mt2g{\mathrm{t}_{2g}}
\def\eg{$\mathrm{e}_{g}$ }

\begin{abstract}
  We investigate the effects of the spin-orbit coupling (SOC) in a three-orbital
  impurity model with Kanamori interaction using the numerical renormalization
  group method. We focus on the impurity occupancy $N_d=2$ relevant to the
  dynamical mean-field theory studies of Hund's metals. Depending on
  the strength
  of SOC $\lambda$ we identify three regimes: usual Hund's impurity
  for $|\lambda|<\lambda_c$, van-Vleck non-magnetic impurity for
  $\lambda > \lambda_c$, and a $J=2$ impurity for $\lambda <
  -\lambda_c$. They all correspond to a Fermi liquid but with very different
  quasiparticle
  phase shifts and different physical properties. 
 The crossover between these regimes is controlled by an emergent
 scale, the orbital Kondo temperature, $\lambda_c =T_K^\mathrm{orb}$
 that drops with increasing interaction strength. This implies that
 oxides with strong electronic correlations are more prone to the
 effects of the spin-orbit coupling.
\end{abstract}

\maketitle


\section{Introduction}

  At energies relevant to solid-state physics, the 
  relativistic effects manifest most prominently as the spin-orbit coupling
  (SOC) term in the Hamiltonian, $H_\mathrm{SOC}=\lambda \mathbf{l}\cdot
  \mathbf{s}$. The alignment of spin and orbital degrees of freedom has several
  effects: at the level of noninteracting electronic structure, it leads to
  the lifting of the band degeneracies and can induce a change of topology in
  the momentum space, a topic widely discussed today; in atoms, the SOC leads
  to the third Hund's rule; in magnetism, it leads to the spin-anisotropies.
  The strength of the SOC increases with atomic number,
  with typical values of 50meV for 3d oxides, 0.1-0.2eV for 4d oxides, and
  about 0.4eV for 5d oxides. Recently the strong effects of the SOC
  have been carefully investigated in 5d oxides, in particular irridates with 5 electrons occupying the
  d-shell, where the energy of the $j=3/2$ states is lowered leading
  to a half-filled $j=1/2$ band and the associated occurence of the Mott
  transition~\cite{Kim2008,Martins2011,Arita2012}.

  The situation in 4d oxides with an intermediate strength of the SOC, notably ruthenates, 
  is more nuanced. On one hand, in band-structure calculations~\cite{Haverkort2008,Liu2008} and
  in spin-sensitive photoemission~\cite{Veenstra2014} the effects of
  the SOC have been
  clearly observed. On the other hand,
  within the dynamical-mean field theory (DMFT) approach the ruthenates have
  been widely and
  succesfully~\cite{Mravlje2011,Stricker2014,Dang2015a,Dang2015b,Gorelov2010,Georges2013}
  discussed as Hund's metals (compounds in which the coherence scale is suppressed by the Hund's
  interaction $J_H$) without taking the effects of $\lambda$ into
  account at all. In
  ruthenates, the SOC is 0.1-0.2eV, which is similar to $J_H$ that is
  about 0.3-0.4eV.  This prompts the question to what 
  scale $T_0$ must one compare the strength of the spin-orbit coupling to
  determine whether its effects are important. Furthermore, for four electrons in
  the t$_{2g}$ shell strong enough SOC leads to a non-magnetic $J=0$ van-Vleck
  insulator regime~\cite{Khaliullin2013, Cao2014, Bhowal2015,
  Meetei2015,Sato2016}, but the threshold SOC strength remains to be
  quantified. Finally, there are also important open qualitative questions. 
  A very recent model DMFT work    found that the SOC  substantially modifies the behavior 
  and leads to an interesting 
  non-Fermi-liquid behavior dubbed the $J$-freezing~\cite{kim2016}. In the absence of SOC, Hund's
  metals are Fermi liquids which follows from the physics of the underlying
  impurity model~\cite{Georges2013}. The three-orbital impurity model with
  SOC has not been explored so far and the nature of its low-energy
  fixed points is unknown.
  
  In the present work we investigate these questions within a
  three-orbital Kanamori impurity model with spin-orbit coupling at
  occupancies $N_d=2$ and $4$ which are relevant to Hund's metals. To solve the model,
  we have implemented a numerical renormalization group (NRG) code
  exploiting the conservation of total angular momentum $J$ to keep the
  computational cost manageable. The ground-state of the impurity
  problem is found to always be a Fermi liquid, but characterizing it
  in terms of the quasiparticle phase-shifts one can distinguish three
  regimes: a Hund's metal for $|\lambda| <
  \lambda_c$, a non-magnetic van-Vleck regime for $\lambda >
  \lambda_c$, and a $j=3/2$ metal for $\lambda < - \lambda_c$.  
  We find that the crossover scale $\lambda_c$ is given not by a bare
  parameter, but rather by an emergent scale: the orbital Kondo
  temperature. We also calculated the impurity spectral function which 
  is found to exhibit characterically different shapes in the three
  regimes. This not only has implications for the physics of oxides
  described within the DMFT but could also be directly observed in the
  tunneling spectra of Hund's
  impurities
  adsorbed on metal surfaces \cite{Khajetoorians2015, Khajetoorians2016, choi16}.


\section{Model}
  We consider a three-orbital impurity model
  $H = H_\mathrm{band} + H_\mathrm{hyb} + H_\mathrm{imp}$
  with the Kanamori Coulomb interaction and the spin-orbit coupling 
  on the impurity site:
\begin{eqnarray}
    \label{eq:Himp} 
    H_{\mathrm{imp}}  = & &\frac{1}{2}(U - 3\Jhm)N_{d}(N_{d}-1) -  \\
      &-& 2\Jhm \mathbf{S}^{2} -
    \frac{\Jhm }{2} \mathbf{L}^{2}+ \epsilon N_d +
    H_\mathrm{ls}.
  \nonumber
\end{eqnarray}
  $U,\Jhm$ 
  are the on-site Hubbard repulsion and Hund's coupling, respectively.
  $N_{d}=\sum_m n_m, \mathbf{S}=\sum_m \mathbf{s}_m, \mathbf{L}=\sum_m
   \mathbf{l}_m$ 
  are impurity total charge, spin and orbital momentum operators, and
  $n_m,\mathbf{s}_m,\mathbf{l}_m$ are the occupancy, spin, and orbital moment in
  orbital $m$. $\epsilon$ sets the occupancy of the impurity. We take the impurity to be coupled to a flat conduction band with
  the density of states \(\rho = 1/2D=1/2\), \( D=1 \) being the half band-width, described by 
  $H_{\mathrm{band}}$.
  %
  $H_{\mathrm{hyb}} = \sum_{k,j,m} V_{k} c_{kjm}^{\dagger}d_{jm}  +
  \mathrm{h.c.}$ is the hybridization.
  The spin-orbit coupling $H_{ls}$ can be written in the spherical orbital
  basis (with $m$=-1,0,1 being the eigenstates of $l_z$ for $l=1$) as
  $H_{ls} = H_{ls}^z + H_{ls}^{xy}$ with
\begin{eqnarray}
    H_{ls}^z = \frac{ \lambda}{2}\sum_{m=-l}^l m(d_{m\uparrow}^\dagger
d_{m\uparrow} - d_{m\downarrow}^\dagger d_{m\downarrow}),\\
    H_{ls}^{xy} =
\frac{\lambda}{2}\sum_{m=-l}^{l-1}\sqrt{(l-m)(l+m+1)}\times\\\nonumber
\times(d_{m+1\downarrow}^\dagger d_{m\uparrow} + d_{m\uparrow}^\dagger
d_{m+1\downarrow}).
\end{eqnarray}
  In this study we consider the model at the impurity occupancies
  $N_d=2$ and $4$. 
  The spin-orbit operator changes sign upon a particle-hole transformation.
  Hence, an impurity occupied by two electrons and SOC $\lambda$ is equivalent to
  an impurity occupied by four electrons (two holes) and SOC $-\lambda$.
  In our discussion we focus on the case with fixed occupancy $N_d = 2$ and
  change the sign of the SOC $\lambda$ to account for the case with $N_d=4$.
  In transition-metal oxides with $t_{2g}$ valence orbitals the
  traditional 3rd Hund's rule is inverted due to the TP correspondence
  which has to do with the fact that the matrix elements of $l=2$ within the
  $t_{2g}$ subspace are the same as these of the $l=1$ operators (within the
  $p$ subspace) but with inverted sign \cite{Sugano1970}.
  $\lambda >0$ that favors small values of $J^2$ is relevant to the
  physics of more than half-filled $t_{2g}$ shell of $d^4$ oxides, such as
  ruthenates, whereas $\lambda <0$ that favors large values of $J^2$ is
  relevant to the less than half-filled $d^2$ oxides, such as
  molybdates,
  as summarized in Table~\ref{tab:convention}.
  
\begin{table}
  \caption{\label{tab:convention}Spin-orbit coupling \(\lambda\), total
  angular momentum \( J \) and occupancy of the $p$ and $t_{2g}$ orbitals.}
    \begin{tabular}{c|c}
    \(\lambda < 0\) & \(\lambda >0\)\\ \hline
    \(J=L+S \)       & \( J=|L-S| \)   \\
    \( p^{4}, d^{2} \) & \( p^{2}, d^{4} \)\\
    molybdates, chromates & ruthenates
    \end{tabular}
\end{table}

\section{Method}
  We solve the impurity problem using numerical renormalization group (NRG)
  solver.~\cite{bulla2008,zitkoNRG} We take into account the conservation of
  charge and total angular momentum to reduce the computational cost. 
  The impact of the exponential growth of the Hilbert space depends on
  the NRG discretization parameter $\Lambda$. In our calculations we used
  $\Lambda =10$.  The effect of quite large Lambda was reduced by using twist
  averaging over $N_z=8$ interleaved discretization grids
  \cite{oliveira1994,campo2005,resolution}.
  In the diagonalization all the states with $E<E_{\mathrm{keep}}=10$ are kept.
  We additionally limit the total number of kept states to 6000 (due to memory
  constraints), a restriction that is more stringent than the former one only in
  the first few iterations. To calculate the spectral functions we use the
  complete Fock space approach~\cite{peters2006}.

\section{Results}

\begin{figure}
    \includegraphics[width=\columnwidth]{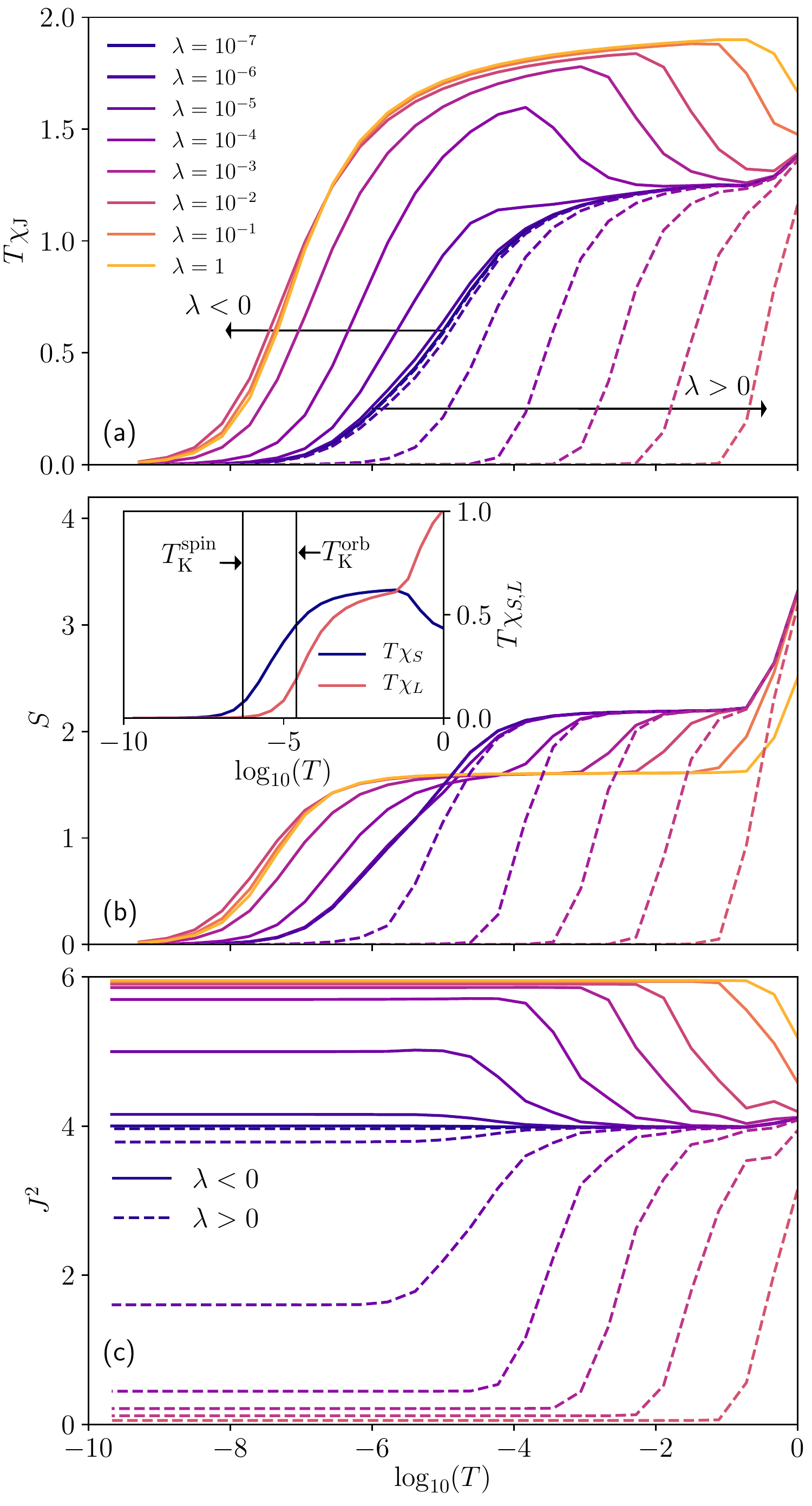}
    \caption{\label{fig:entropy} 
    (a) Effective local moment. Inset: effective spin and orbit moment
    for \( \lambda=0 \).
    (b) Impurity entropy. 
    (c) Expectation value \( \langle J^{2}\rangle \).
    Solid (dashed) lines denote results for $\lambda < 0$ ($\lambda > 0$).
    }
\end{figure}

  We first discuss the influence of the spin-orbit coupling on the thermodynamic
  expectation values. In Fig.~\ref{fig:entropy}(a) we display the
  temperature evolution of the effective local moment evaluated as \(
  \chi_J T\), where $\chi_J$ is the impurity contribution to the total
  angular moment susceptibility. Panel (b) shows the impurity contribution
  to entropy and panel (c) shows the total angular momentum at the impurity
  site $\langle J^2 \rangle$. Throughout the text the parameters are 
  $U = 3.2$, $\Jhm=0.4$, and $\Gamma = \pi \rho_0 V^2  = 0.05$.
  
  Consider the $\lambda=0$ case first. With decreasing temperature
  $T$, after the charge fluctuations are
  frozen out (above temperatures shown) the model enters into the
  local-moment 
  regime with a 9-fold degenerate $L=1, S=1$ multiplet characterized by
  a plateau in the entropy at a value of $\log(9)$ and with $\chi_J T
  \approx \langle J^2 \rangle/3$ where $\langle J^2\rangle = \langle L^2
  \rangle + \langle S^2 \rangle = 4$. (This result follows also in the $J$
  basis, \( \langle J^{2}\rangle = (6\cdot 5 + 2\cdot 3 + 0\cdot 1)/15=4
  \).) On further lowering the temperature, the local moment is progressively screened
  and becomes small below a low (Kondo) temperature. 

The inset to
  Fig.~\ref{fig:entropy} separately displays for the $\lambda=0$ case
  the local spin moment $\chi_S T$ and the orbital angular moment
  $\chi_L T$ evaluated respectively from spin and orbital susceptibilities.
  As discussed in earlier work\cite{Yin2011, Aron2015, horvat2016} the screening of the spin
  moment occurs at a lower temperature $T_K^\mathrm{spin}$ than the one for the orbital
  moment $T_K^\mathrm{orb}$ (the two Kondo temperatures are indicated by the
  two vertical lines and differ by about an order of magnitude). The initial
  drop of $\chi_J T$ and the associated suppression of the impurity contribution
  to entropy, seen in Fig.~\ref{fig:entropy}(b), thus comes mainly from the
  quenching of the orbital degrees of freedom.

  Turning on the spin-orbit interaction has a strong effect with
  markedly different behavior in the cases of the positive (dashed lines) and negative (plain
  lines) values of $\lambda$. For $\lambda < 0$, the SOC tends to align 
  the spin and the orbital moment to a
  state of a large total angular moment \( J=L+S \). The expectation
  value of \( J^{2} \) tends to $6=2(2+1)$, and $\chi_J T$
  approaches 2 in the high-temperature local-moment regime when the
  temperature is lowered to $T<|\lambda|$. On cooling down further, the
  local moment is screened; the corresponding Kondo temperature is
  found to diminish as $|\lambda|$ increases. 
  In the temperature window $T_K < T <|\lambda|$, the
  impurity contribution to entropy shows a clear plateau at $\log 5$
  revealing the degeneracy of the $J=2$ local moment. For intermediate
  strenghts of $|\lambda|$ one sees first a plateau at the $\log 9$ S=L=1
  manifold and then a crossover to the $\log 5$ value when the
  temperature drops to a value $T<|\lambda|$. 

  For $\lambda>0$, the SOC tends to anti-align the spin and the
  orbital moments which leads to a non-degenerate \( J=0 \) atomic
  ground state. This is a peculiar ``no-impurity'' regime of an
  impurity problem: there are no internal degrees of freedom at the
  impurity and the conduction electrons experience only potential
  scattering.  It turns out that this atomic consideration describes
  the numerical results well, provided that $\lambda$ is significantly
  larger than the Kondo temperature itself.  
  
  We now turn to $\langle J^2 \rangle$ shown in Fig.~1(c). For large
  $\lambda>0$, as soon as temperature drops below $\lambda$, $\langle
  J^2 \rangle$ rapidly approaches a very small value. At the same
  temperature, the entropy and $\chi T$ also drop rapidly to 0, as
  seen in panels (b) and (a). Notice the distinct behavior of $\langle
  J^2 \rangle$ and $\chi_J T$ for positive and negative $\lambda$. In
  the latter case, one has a large atomic moment that is screened by
  the quantum fluctuations on cooling below the Kondo temperature
  whereas in the former case the atomic moment is not present to start
  with. Or, thinking in terms of the temperature dependence, for
  $\lambda>0$ the
  suppression of $\chi_J T$ and $\langle J^2 \rangle$ occurs at the
  same scale $T\approx \lambda$: the process is atomic and does not
  involve the conduction electrons. 

\newcommand{\expv}[1]{\langle #1 \rangle}

\begin{figure}
    \includegraphics[width=\columnwidth]{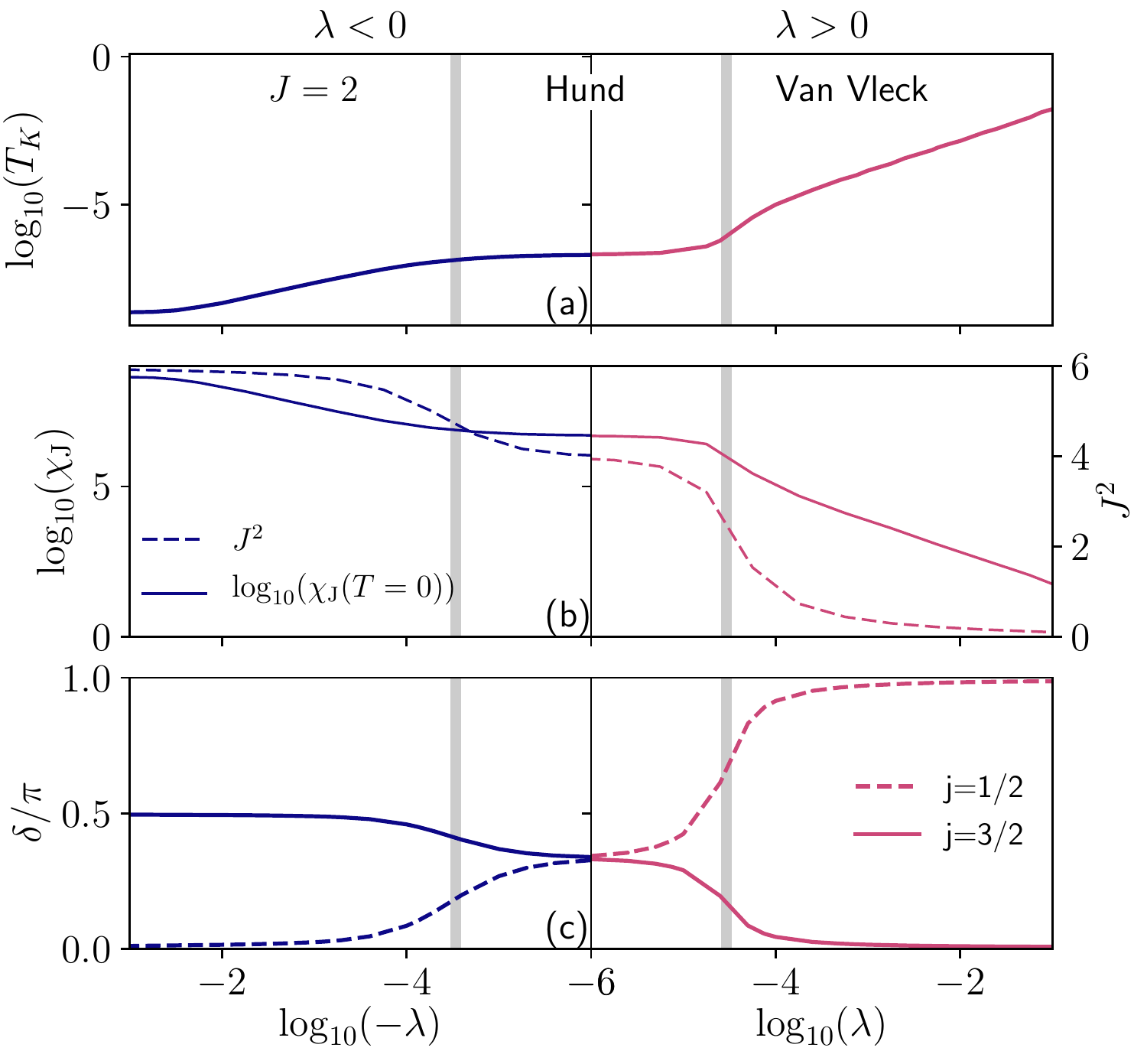}
    \caption{\label{fig:J2-chi}
    (a) Kondo temperature.
    (b) Zero-temperature total-angular-momentum susceptibility
    \(\chi_J\) (full line) and expectation value
    \( \langle J^{2}\rangle \) (dashed line).
    (c) Quasiparticle phase shifts.
    }
\end{figure}

  The evolution of key quantities as a function of $\lambda$, shown in
  Fig.~\ref{fig:J2-chi}, serves to delineate the different regimes.
  In Fig.~\ref{fig:J2-chi}(a) we present $T_K$ defined as the
  temperature at which $\chi_J T$ drops below 0.07. For $\lambda < 0$
  (left from centre) we observe a reduction of $T_K$ with increasing
  $|\lambda|$.  The lowering of the Kondo temperature is due to the
  formation of a larger moment combined with the splitting of the
  multiplets that leads to a suppression of the Kondo coupling
  strength.  
  From the Schrieffer-Wolff transformation 
  one finds that in the limit $|\lambda|\ll J_H,U$ the Kondo couplings
  are proportional to 
  $1/(c_0 - \lambda), c_0>0$.
  For $\lambda > 0$ (right from centre) the opposite behavior is seen.
  For large $\lambda$, $T_K \approx \lambda$\footnote{One needs to be
  careful in interpreting $T_K$ as a Kondo temperature in this regime
  as the quenching of the moment is not due to conduction electrons
  but is rather atomic.}. A very similar dependence on $\lambda$ is
  found also in the zero-temperature total angular momentum susceptibility $\chi_J$
  that is shown in Fig.~\ref{fig:J2-chi}(b) along the expectation
  value $\expv{J^2}$. 
  The evolution of $T_K$ as a function of the SOC $\lambda$ 
  is indicative of the one found for the inverse of mass
  enhancement $Z$, $T_K \sim Z=(1-\partial \mathrm{Re}[
\Sigma(\omega)]/\partial \omega)^{-1}|_{\omega
\rightarrow 0}$
  (in our results $Z$ and $T_K$ are monotonously related 
    with a bit weaker dependence of $Z$ on $\lambda$).

  For any value of $\lambda$ and irrespective of its sign, the local moment
  is completely screened yielding a regular Fermi liquid behavior at low
  temperatures. Hence, one can map the low-energy excitation spectrum to that
  of a non-interacting resonant-level model and parametrize it in terms of
  the quasiparticle scattering phase shifts \(\delta\). In
  Fig.~\ref{fig:J2-chi}(c) we present the phase shifts corresponding to the
  \( j=1/2, j=3/2 \) excitations as a function of \(\lambda\).
  They are associated through the Friedel sum-rule to the occupancies
  of the corresponding resonant levels
\begin{equation}
    \delta_j = \pi n_j /(2j +1).
\end{equation}
  The regime of large negative $\lambda$ corresponds to a half-filled 4-fold
  degenerate $j=3/2$ resonant level and empty $j=1/2$ state. Conversely, at
  large positive $\lambda$ the $j=3/2$ states are emptied out and one is
  left with the completely occupied $j=1/2$ states.

  The SOC needs to be increased in absolute value above some critical
  value \( \lambda_c \) in order to have any observable effect. 
  This value, $\approx 10^{-5}$ for parameters used in this work, does not correspond to any of the bare scales of the problem but is associated to a low temperature emergent scale,  the Kondo temperature. The qualitative explanation of this surprising finding is as follows: Considering the
  problem from the perspective of the renormalization group and
  progressively lowering the temperature, it is clear that if the
  moments are screened on a scale higher than $\lambda$, the SOC has
  nothing to act on~\cite{Mravlje2016}. In the case of the Hund's
  metal, where the orbital moments are screened first, one can be more
  precise: $\lambda_c$ is determined by the onset of screening of the
  total angular momentum, which corresponds to the orbital
  Kondo temperature, which is the higher of the two Kondo screening
  scales. Thus $\lambda_c \approx T_{K}^\mathrm{orb}$. 
  This is most easily seen by inspecting the phase shifts that evolve
  by a half of the change to the final value at $\lambda=T_K^\mathrm{orb}$.



\begin{figure}
    \includegraphics[width=\columnwidth]{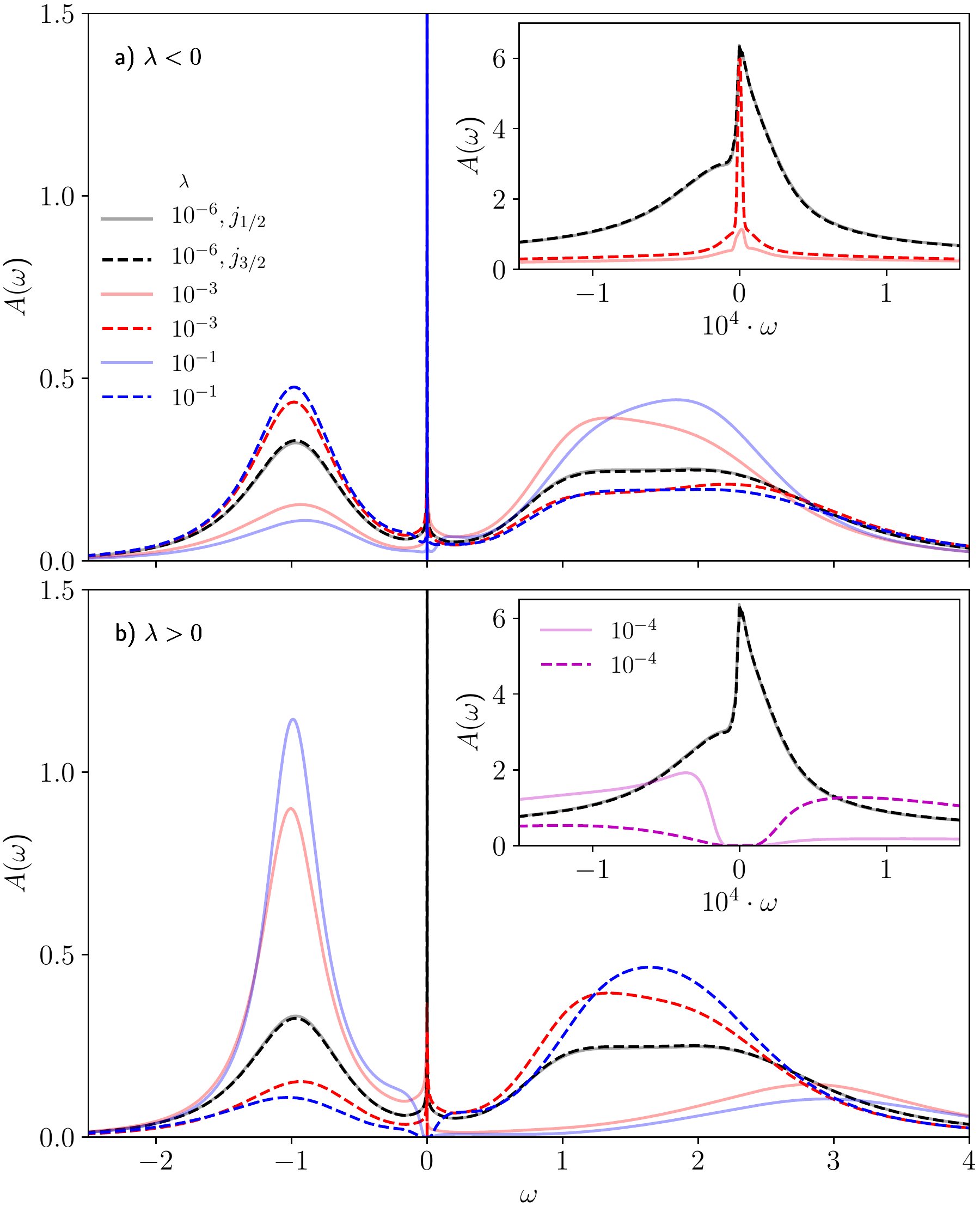}
    \caption{\label{fig:spect-fun} 
    Impurity spectral functions in the presence of the spin-orbit
    coupling for $j=1/2$
  (full lines) and $j=3/2$ (dashed) excitations.
  The inset are close-ups on the low-energy region. 
}
\end{figure}

  We now turn to the impurity spectral functions shown in 
  Fig.~\ref{fig:spect-fun}. 
  For $\lambda=0$, the $j=1/2$ and $j=3/2$ excitations are degenerate, hence
  the corresponding spectral functions remain almost the same as long as
  $|\lambda| < \lambda_c$. One can resolve a lower Hubbard band, a broad
  upper Hubbard band (that contains excitations to different multiplets of
  half-filled impurity), and a Kondo resonance. The latter has a
  characteristic asymmetric shape that is a fingerprint of the Hund's
  metal\cite{Stadler2015} (see also \cite{Wadati2014}).

  When the magnitude of the spin-orbit coupling is increased, the
  degeneracy between the $j=1/2$ and the $j=3/2$ states is lost. In
  $\lambda < 0$ regime the $j=1/2$ states are pushed to positive
  energies (with a low value of the spectral function at $\omega =0$,
  the small hump seen in the inset of Fig.~\ref{fig:spect-fun} (a) in
  $j=1/2$ spectral function for $\lambda=10^{-3}$ is a discretization
  artefact); at large $|\lambda|$ this leads to a half-filled $j=3/2$
  level with nearly symetrical lower and upper Hubbard bands. The
  suppression of the Kondo temperature during the crossover to 
  $J=2$ regime is reflected in the narrowing of the Kondo resonance.

  Quite different physics occurs for \(\lambda >0\). 
  The \( j=1/2 \) spectral weight is pushed to negative energies and the
  \( j=3/2 \) spectral weight to positive energies. The Kondo resonance
  is split by the SOC, see the inset, with the splitting being of order
  $\lambda$.  The spectral weight at $\omega=0$ is small, an indication
  of the insulating-like behavior. This regime hence corresponds to that
  of a ``non-magnetic van-Vleck insulator''.  

\section{Discussion} 
  It is important to note that in the context of real materials, the orbital
  Kondo temperature can be substantially higher than the low-temperature
  Fermi liquid coherence scale. In ruthenates, the orbital moments are
  screened at 1000K (0.1eV)~\cite{Mravlje2016} which is similar to the
  estimated values of the spin-orbit coupling (0.2eV). This indicates
  that the ruthenates
  are within the Hund's metal regime and explains why calculations
  neglecting the SOC obtain reasonable results.
 
  Our results suggest that the spin-orbit interaction could diminish the
  coherence scale in $d^2$ systems. Molybdates have such occupancy but are
  characterized by a coherent behavior with high Kondo
  temperature~\cite{Wadati2014} because they are far from van-Hove
  singularity and hence are not appreciably affected by the SOC. Perhaps
  chromates~\cite{Medici2011} realize a $J=2$ metal (in spite of the smaller
  SOC) due to the strong correlations and hence small orbital Kondo
  temperature.

\section{Conclusion}
  In summary, we investigated the effects of the SOC in a
  three-orbital impurity occupied by two electrons/holes
  that we have solved with the NRG. In $d^2$ systems the SOC leads to a
  crossover from a Hund's impurity with a distinct behavior of spin and
  orbital moments to a $J=2$ impurity with a suppressed Kondo
  temperature. In $d^4$ systems the crossover is to the non-magnetic
  regime with no local moment ($J=0$) instead. The spectral functions
  in different regimes are characteristically different. The SOC
  becomes effective once it exceeds the emergent
  low-energy scale, the orbital Kondo temperature.  Besides implication it has for oxides\footnote{See also a very recent preprint, M. Kim {\it et al.} arXiv:1707.02462 for a DMFT study that finds behavior fully consistent with the one found in the impurity model here}, this finding 
 could be  tested in an STM experiment where one
  would control the Kondo temperature and observe the change in the
  shape of the spectral function as the orbital Kondo temperature
  falls below the SOC strength. 

{\it Acknowledgments}
We thank M. Aichhorn, A. Georges, A.J. Kim, M. Kim, and R. Triebl for useful discussions.  We acknowledge the support of the Slovenian Research Agency (ARRS)
under P1-0044.

\end{document}